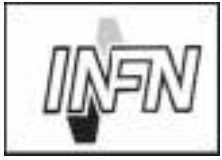

ISTITUTO NAZIONALE DI FISICA NUCLEARE

Sezione di Firenze

# Antimatter, just 'semantic' question

Piero Spillantini[1]

[1) I.N.F.N. e Univ. di Firenze, I-50121 Firenze, Italy

**Abstract**

In the attempt of including leptons in the scheme of a matter-antimatter asymmetry originated from CP violation in the primordial Universe, it is proposed of re-christening $e^-$, *muon*$^-$, *tau*$^-$ and their associated neutrinos as 'antiparticles' and $e^+$, *muon*$^+$, *tau*$^+$ and their associated antineutrinos as 'particles'.

# 1 Introduction

In the field of Cosmology most of the scientificfic works concerning the question of the matter-antimatter asymmetry of the Universe deal with the baryonic matter without mentioning the role and the evolution of the leptonic matter. The observation of this lack gave rise to the attempt outlined in this letter of including also the leptons in the simple scheme of a matter-antimatter asymmetry originated from CP violation in the breaking of the supersymmetric state of the primordial Universe.

A further reason, much more speculative and 'academic', for such an attempt could be found in the dissatisfaction for the final fate of an 'all-matter' Universe that the possible decay of the proton drives to an 'all-antimatter' Universe. It must anyway be underlined that this argument, even if it could have an aesthetic merit, cannot at all be regarded as a scientific one: the mean life of the proton is in any case so long, exceeding by tens of order of magnitude the present life time of the Universe, that any extrapolation of the evolution of the Universe on such a long time scale is a nonsense.

# 2 The particle-antiparticle dichotomy

Let us comment on the current use of the dichotomous pair of words particle-antiparticle and matter-antimatter.

In Elementary Particle Physics the word 'antiparticle' defines an 'entity' that has the same properties of the corresponding entity defined as 'particle', but opposite characteristic charges of its fundamental interactions. We know, both from theory and from experiments, that particles and antiparticles are produced in pairs. This rigorous result of Elementary Particle Physics implies that we can assume that, just after the Big Bang, the content of the Universe in particles and antiparticles was identical. In fact the current cosmological theories resulting either in a baryon symmetric Universe or in a baryon asymmetric one imply that the difference of particle and antiparticle content of the Universe must be zero at the very beginning.

On the other hand cosmological theories must introduce a relevant violation of CP symmetry during the phase transition from a highly symmetric state to a less symmetric one, either for constructing baryon or antibaryon dominance in separated regions of the Universe, or for constructing an asymmetric Universe, fully dominated by baryons. When applied to supersymmetric theories, the CP symmetry violation gives different rates for the decays of the parent heavy bosons in channels which do not conserve the baryon number. In a simple 'symbolic' formulation the decays that would occur with different branching ratios can be expressed by the decay of the heavy boson in quarks and leptons and of the heavy antiboson in antiquarks and antileptons. As the quarks are the constituents of the nucleons (i.e. of the stable baryons), it means that the final result of the heavy boson decays must be the stable positively charged baryons, the protons. Since charge must be conserved, to these positively charged baryons must correspond negatively charged leptons, the electrons.

Clearly in this scheme, if the CP violation mechanism is at work, the total difference between the numbers of particles and of antiparticles is altered. But this alteration is both unnecessary and 'anaesthetic'. It is unnecessary because the 'symbolic' parent bosons decays could be expressed by the decay of the heavy bosons in quarks and antileptons and that of the heavy antibosons in antiquarks and leptons. By this re-writing the final stable positively charged baryons, the protons, are generated coupled to the negatively charged antileptons, the electrons, with no consequences in the whole scheme of the elementary particle theories. For symmetry reasons the final products of the heavy antiboson decays could be considered as antibaryons (antiprotons at the end) and leptons (positrons at the end).

## 3  One unique primordial boson?

As a consequence of the above re-writing, also the split of the primordial parents in bosons and antibosons becomes superfluous: the antiquarks-leptons and quarks-antileptons final states can both proceed from a common parent heavy boson. We could indicate it by the capital letter Z in order to mean its 'zero' particle number, possibly by using a suitable graphic that could avoid confusion with the $Z°$ boson mediating the weak interactions. Another, perhaps more appropriate, indication could be by the capital letter P: it avoids any confusion with other symbols already in use for particles (remember that the capital 'P' indicating the Pomeron has a somewhat special graphic) and can be understood as 'Primeval' or 'Primary' or simply 'Parent' boson.

All this is self-consistent, and has the advantage of saving the symmetry between the particle and antiparticle content of the Universe, or at least of conserving the difference between its content in particles and its content in antiparticles, also in a framework where baryonic and leptonic numbers can change.

## 4  Re-christening of leptons and antileptons

This question is not raised in the framework of the standard model, where baryons and leptons are separately conserved 'worlds'. And in fact, if $e^-$, *muon*$^-$, *tau*$^-$, and their neutrinos, are defined as antiparticles and $e^+$, *muon*$^+$, *tau*$^+$, and their antineutrinos, as particles the general scheme of particle interactions remains unchanged, and the definition of 'particle' and 'antiparticle' can be considered just a semantic question.

The question of the conservation of the particle-antiparticle difference in the Universe arises in the framework of theories allowing the baryon-lepton transition. In these theories the 'useless' violation of this difference can be avoided by simply defining as antiparticles all the particles until now known as 'leptons'. A straightforward simple example is the supposed decay of the proton in one $\pi°$ and one $e^+$: the renaming of the $e^+$ as particle avoids the violation by two units of the particle-antiparticle difference. The most remarkable consequence of this is a 'highly academic' one, a nonsense from a scientificfic point of view, as pointed out above, but it satisfies the aesthetics: after a time as long as the mean lifetime of the proton the symmetry between particles and antiparticles is maintained, what would not be the case in the definition of particles and antiparticles for leptons and antileptons.

As a conclusion, in order to simplify and rationalize the overall scheme, the $e^+$, *muon$^+$*, *tau$^+$* and their associated antineutrinos should be 're-christened' as 'particles', and the $e^-$, *muon$^-$*, *tau$^-$* and their associated neutrinos as 'antiparticles'.

This re-christening could leave unchanged the definition of matter and antimatter, with the stable matter constituted by protons, neutrons and electrons, i.e. the semantics could be:

$$\text{Matter} = \text{Quarks} + \text{Antileptons}$$
$$\text{Antimatter} = \text{Antiquarks} + \text{Leptons}$$

## 5 Conclusion

The re-christening of $e^-$, *muon$^-$*, *tau$^-$* and their associated neutrinos as 'antiparticles' and of $e^+$, *muon$^+$*, *tau$^+$* and their associated antineutrinos as 'particles' could be regarded as of negligible importance, being substantially a question of semantic definition, not contributing to the progress of our understanding of the fundamental laws of the physical world.

However such re-christening gets rid of the above mentioned aesthetic dissatisfaction. Aesthetics has always been a powerful tool in physics, and also the semantics could result not disjointed from the essence which it voices, and therefore deserves to be settled.